\documentstyle[11pt,moriond,epsfig]{article}

\def\Journal#1#2#3#4{{#1} {\bf #2}, #3 (#4)}

\def\NPA{{\em Nucl. Phys.} A}
\def\NPB{{\em Nucl. Phys.} B}
\def\PLB{{\em Phys. Lett.}  B}
\def\PRL{\em Phys. Rev. Lett.}
\def\PRC{{\em Phys. Rev.} C}
\def\PRD{{\em Phys. Rev.} D}

\def\be{\begin{equation}}
\def\ee{\end{equation}}
\def\bea{\begin{eqnarray}}
\def\eea{\end{eqnarray}}

\begin{document}
\vspace*{4cm}
\title{ALICE, the Heavy Ion Experiment at LHC \footnote{Talk given at 
the XXXVIth RENCONTRES DE MORIOND QCD High Energy Hadronic Interactions, Les Arcs, Savoie, France, March 17-24, 2001 }
}

\author{Gin\'es MARTINEZ for the ALICE collaboration }

\address{SUBATECH, BP 20722 44307-Nantes-Cedex, France  \\
e-mail : martinez@in2p3.fr}

\maketitle\abstracts{
ALICE, A Large Ion Collider Experiment, is the future Large Hadron Collider (LHC) 
experiment at CERN devoted to the physics of Quantum Chromo-thermo-dynamics.
Relativistic Heavy-Ion Collisions (HIC) at LHC 
aim at the production of a plasma of quarks and gluons (QGP).
This plasma is expected to be much hotter, bigger and longer  
than in previous HIC experiments at lower center-of-mass energies.
In the ALICE experiment, the ephemeral QGP created during the first stages of the HIC will be studied by 
the concomitant detection of most  of the probes of high temperature strongly interacting matter.
}

\section*{Why a Heavy Ion Experiment at LHC?}

In our world, quarks and gluons are confined in colorless states called hadrons, 
which define the degrees of freedom of low temperature nuclear matter.
However, at high energy densities, lattice calculations of the theory of strong interactions (QCD) 
predicted the deconfinement of quarks and gluons, which become the {\it right} 
degrees of freedom of a new state of matter \cite{MacL81}. 
Recent calculations \cite{Kars00} show that for a system at zero baryon density,  
the transition temperature between the hadron and parton phases is around $T_c=175$ MeV and 
the critical energy density  $\epsilon_c=700$ MeV/fm$^3$.
In addition, this transition is expected to be accompanied by the restoration of the chiral symmetry of the QCD interaction, 
which is spontaneously broken at low energies \cite{Kars98}.
The study in the laboratory of both transitions as well as the properties  of the QGP   
are the major goals of the future ALICE experiment at LHC (CERN) \cite{ALICE}.

Relativistic heavy-ion collisions (HIC) are a unique experimental tool to investigate the 
QGP phase diagram in the laboratory.
During the last 20 years, HIC have been largely exploited up to SpS energies ($\sqrt{s} \sim 20A$ GeV) \cite{QM99}.
However a detailed study of the phase transition and/or of the QGP has not been possible yet, 
due to the complexity of the HIC dynamics and the difficulty to establish the links 
between the observed probes and the QGP properties.
As a matter of fact, at SpS energies:
i) perturbative QCD calculations  can not be exploited to describe the dynamics of the HIC  where soft processes are dominant;
ii) the size of the hot strongly interacting system is given by the available number of valence quarks;
iii) the QGP, if formed, has temperatures close to the critical one, which renders 
difficult to disentagle between the QGP and the hot hadron gas;
iv) nuclear stopping is too strong to create a bulk matter with vanishing baryon density.

HIC at the LHC ($\sqrt{s}=5.5A$ TeV for Pb+Pb collisions)  
will open new perspectives in the study of QGP properties in the laboratory.
The first stage of the HIC at these energies can be described by the saturation scenario \cite{MacL94} 
where a large number of sea gluons, characterized by small values of x=$2p_T/\sqrt{s}$, 
will be freed in the beginning of the collision, leading to the formation of large system of 
interacting partons with zero baryon density. 
Up to 8000 gluons in the early stage of the collisions are predicted \cite{Venu01}. 
Such a system will rapidly evolve towards equilibrium in a process that can be described by the asymptotically free field theory of QCD \cite{Baie01}. 
After equilibration, a very hot plasma, $\epsilon\sim 25\epsilon_c$, will be formed, 
in a volume 10 times larger than at SpS energies and for longer life-times, up to 10 fm/c.
In addition, particle production will be dominated by hard and semi-hard processes which can be theoretically described 
by perturbative QCD \cite{Wang91}.
In summary, the remarkable increase of the energy and the size of the strongly interacting system,
will allow for an easier connection between the experimental probes and the properties 
of the zero-baryon density QGP.

\section*{The probes}
The experience acquired during the last 20 years of heavy-ion physics in the relativistic regime has 
shown the necessity to measure most of the probes of the reaction dynamics, 
from hard pre-equilibrum processes and QGP formation observables, until the freeze-out of the expanding hadron gas.
A coherent explanation of the full set of observables will be the only way to 
study the properties of the ephemeral QGP. 
In this context, ALICE strategy is to study concurrently all the probes in the same experiment together with global
information of the event topology: particle multiplicity, forward energy, transverse energy. 
Final states probes like particle multiplicities, hadron $p_T$ distributions, particle ratios, strangeness production  
will tell us about the conditions of the phase transition and the dynamical evolution of the expanding hadron gas. 
Penetrating probes, like real and virtual photon production, charmonium suppression 
and in-medium light vector meson properties  
will provide us with information about the QGP phase during the first stages of the HIC.
In addition, the study of the QGP at LHC energies will be enriched by exploiting new  
probes which can be efficiently studied in this energy regime:
\begin{itemize}
\item Hard processes leading to energetic partons will provide information about the QGP 
as they interact with the dense surrounding partonic medium \cite{Wied00}. 
Parton energy losses in the QGP will modify the hadronization process 
of the produced partons, leading to a final-state suppression of high p$_T$ hadrons (jet-quenching).
\item The Debye screening of bottonium bound states in QGP will be studied. 
Suppression of the Upsilon familly production will be measured due to its sensitivity to the
density of color charges in the medium \cite{Satz86}.
\item Open charm and open beauty production will be also accessible at LHC \cite{Wong98}. 
In particular, possible enhancement of the open charm production will probe  
the equilibration process of the large partonic system with an initial temperature 
close to the charm quark mass scale \cite{MacL01}. 
\item Finally, the huge particle multiplicity at LHC 
will allow the measurement of a large number of observables in an event-by-event basis, 
increasing the sensitivity to non-statistical fluctuations predicted to occur in a phase transition scenario.
\end{itemize}

\section*{Building ALICE}
Heavy-ion integrated luminosity at the LHC will be limited by the short-beam time as a consequence 
of the large cross-sections for electromagnetic processes.
For the lead beam, the luminosity is limited to 10$^{27}$ cm$^2$s$^{-1}$, leading to a  
beam life-time \cite{Mors01} of 5 h \footnote{3h life-time if two experiments participate in the LHC heavy ion run.}.
This corresponds to less than 8000 minimum bias interactions per second.
Only 5-10\% of those interactions will correspond to the most central collisions.
Each Pb+Pb central collision will produce large particle multiplicities,
up to 8000 charged  particles per rapidity unit are predicted by theoretical  models \cite{Arme00}. 
Therefore low interaction rates and large particle multiplicities 
are the main design considerations of the ALICE experiment.
The main characteristics of ALICE (see Fig.1) are \cite{TRDS}:

{\it High density particle tracking.}
The Time Projection Chamber (TPC) is the main element of the ALICE tracking system.
Two tracks separation, energy loss resolution better than 10\% 
and large acceptance, $|\eta|<0.9$, define the geometrical parameters of this huge TPC: 
inner radius around 80 cm, outer radius 2.5 m and  overall length of 5 m.
With  about 5700,000 channels the TPC information is the largest element of the ALICE DAQ event with 
around 60 MBytes per central Pb+Pb collision.

{\it Particle Identification.}
ALICE PID system consists of:
i) A large acceptance hadron identification system. 
Hadrons are identified over the TPC geometrical acceptance with a time-of-flight detector (TOF)
in the intermediate p$_T$ range below 2.5 GeV/c. 
TOF basic element consists of Multigap Resistive Plate Chambers placed at 3.7 m from the interaction point,
achieving less than 100 ps time resolution.
ii) A small acceptance high-p$_T$ (p$_T<$5 GeV/c) hadron identifier (HMPID),  consisting of 7 RICH modules
covering around 1 unit of pseudo rapidity.  
Cherenkov photons produced in the $C_6F_{14}$ radiator are converted by a $CsI$ photo-cathode coupled to a MWPC. 
iii) A large acceptance electron identifier (TRD) for p$_T>1$ GeV/c based on transition radiation technique:
X-rays produced in the radiator are detected by a Time Expansion Chamber (TEC) 
operating with a xenon-based gas mixture. 
Up to 6 Radiator-TEC layers  are needed to reach the required pion rejection of $10^{-3}$, for an electron efficiency larger than 90\%. 
In addition, PID of very low p$_T$ tracks (p$_T\leq$ 600 MeV/c) will be performed 
by energy loss measurements, dE/dx, in the TPC and the Internal Tracking System (ITS).

{\it Secondary vertex detection.} 
The ITS of ALICE allows for a determination of secondary vertex from charm and hyperon decays and  the measurement of the primary vertex 
with very good spatial resolution.
It consists of six cylindrical (inner radius of 4 cm and outer radius of 44 cm)  layers of silicon detectors: 
2 silicon pixel, 2 silicon drift and 2 silicon strip layers, over $|\eta|<0.9$ acceptance window.
Combination of information from the TPC and ITS allows for an improvement of the momentum resolution 
of the tracking system. A resolution better than 2.5\% at momenta about 4 GeV/c is obtained.

{\it Photon detection.} Direct and decay photons are detected by the Photon Spectrometer (PHOS). 
The large particle multiplicity demands a Moli\`ere-radius of the calorimeter crystals  
to be as small as possible, and a good energy resolution 
for the measurement of neutral mesons by invariant-mass analysis of photon-pairs
in a huge combinatorial background.
The geometrical acceptance is fixed by the detection of neutral-mesons in the $p_T$ range from 1 to 5 GeV/c.
17920 crystals ($2.2\times2.2\times18.0$ cm$^3$) of PbWO$_4$ scintillator grouped in 5 modules at 4.6 m from the interaction vertex
fulfill these requirements.
  
{\it Muon detection.} The muonic channel will be studied in the forward muon spectrometer.
In this pseudo-rapidity domain, 2.5$< \eta <$4.0, 
most of the hadrons, photons and electrons from the vertex are stopped by a composite absorber (see Fig.1).
Muon trajectories along the dipole magnet will be measured by 10 plane tracking system consisting of Cathode Pad Chambers. 
A mass resolution better than 100 MeV/c$^2$ is required in order to be able to separate the different resonance states 
of each quarkonium family. Muon trigger and identification will be ensured 
by a passive filter wall followed by 4 plane RPC chambers operating in streamer mode. 
\begin{figure}
\begin{center}
\psfig{figure=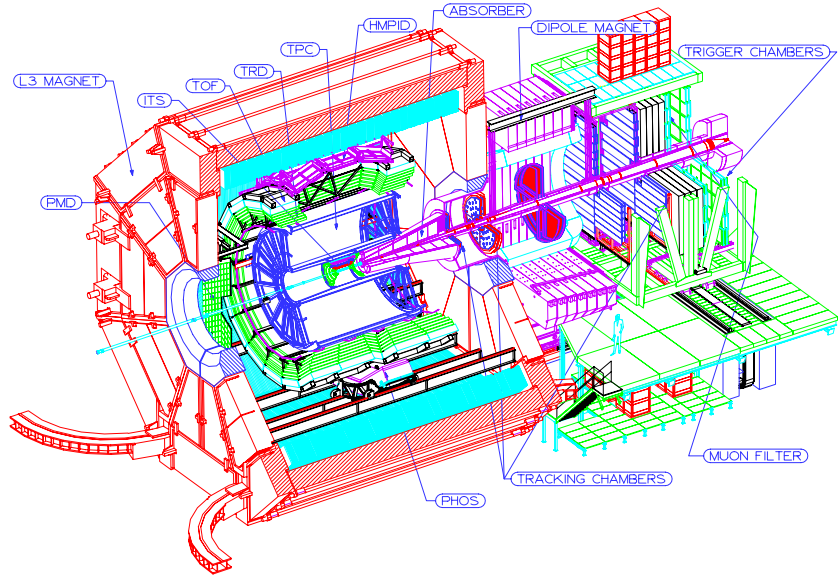,width=11cm}
\caption{Layout of the ALICE detector.
\label{fig:layout}}
\end{center}
\end{figure}

\section*{Acknowledgments}
I would like to thank the {\it Rencontres de Moriond} organization for this exciting conference,
David d'Enterria, Joaqu\'{\i}n Garc\'{\i}a, Guy Paic, 
Jurgen Schukraft and Yves Schutz for fruitful discussions,
to express my gratitude to the colleagues working in the preparation of ALICE and 
to the ``Conseil R\'egional des Pays de la Loire'', France.

\end{document}